\documentclass[a4paper]{article}
\begin{document}

\centerline{\huge \bf All is $\Psi$}
\vspace{20pt}

\centerline{ \bf Lev Vaidman}

\centerline{Raymond and Beverly Sackler School of Physics and Astronomy}
 \centerline{Tel-Aviv University, Tel-Aviv 69978, Israel}

\vspace{5pt}

\begin{abstract}
It is argued that  standard quantum theory without collapse provides a satisfactory explanation of everything we experience in this and  in numerous parallel worlds. The only fundamental ontology is the universal wave function evolving in a deterministic way without action at a distance.
\end{abstract}

\vspace{3pt}

\section{Introduction}

This is a contribution to the 3rd International Symposium about Quantum Mechanics based on a ``Deeper Level Theory'', but the message I want to present is that there is no need for a deeper theory. In my view bare quantum mechanics provides a good explanation of all that we see around us. In fact, it seems that this series of the Symposia leads in this direction. The  quote from the  objectives of the EmerQuM (2011) reads:
\begin{quote}
The theme of ``emergent quantum mechanics'' is, we believe, an appropriate present-day topic, which can both serve as ... a means to point towards promising future directions in physics. We intend to bring together many of those physicists who are interested in or work on attempts to understanding quantum mechanics as emerging from a suitable classical (or, more generally: deeper level) physics...
\end{quote}
The objectives of EmQM13 (2013) were not very different:
\begin{quote}
The symposium invites the open exploration of an emergent quantum mechanics, a possible ``deeper level theory'' that interconnects three fields of knowledge: emergence, the quantum, and information. Could there appear a revised image of physical reality from recognizing new links between emergence, the quantum, and information? Could a novel synthesis pave the way towards a 21st century, ``super-classical'' physics? ...
\end{quote}
However, the first objective  of EmQM15 (2015) is changed:
\begin{quote}
The symposium invites the open exploration of the quantum state as a reality. The resurgence of interest in ontological quantum theory, including both deterministic and indeterministic approaches, challenges long held assumptions...
\end{quote}

I believe the quantum state is a reality. It is the reality. The only fundamental physical ontology is the quantum wave function.

Please allow me a sociology of science speculation for why this opinion is not in the consensus today. At the birth of quantum mechanics there was at least a decade of quantum mechanics without the wave function and the Scr\"odinger equation of its evolution. During this time Bohr with his charismatic behaviour convinced the physics community that the reality behind quantum mechanics cannot be grasped by human beings. A half century later,  physicists working on the foundations of quantum mechanics expressed growing dissatisfaction with Bohr's view, they  followed Bell \cite{BellBeab} in search of ``local beables''. The ingenious model of Spekkens \cite{Spek} gave a hope that our familiar everyday reality will  emerge from a ``deeper level theory''.

However, the research triggered by these developments cooled down the expectations. It suggested that whatever the ``deeper theory'' is, it cannot be simpler than the standard quantum theory \cite{Leifer} and it is strongly suggested that the quantum wave function is ``ontic'' \cite{PBR,Hardy}, i.e., it is the ontology of quantum mechanics.

Moreover, it was claimed that just a ``free will'' assumption is enough for a simple proof of an ontic nature of the quantum wave function \cite{CR}. In my view, however, this argument is a metaphysical error. In science we {\it assume} ``free will'' of an external agent who tests  theories we construct, so nothing can be derived from the ``free will'' assumption.

 It is considered ``politically incorrect'' to claim that physicists, at large, understand Nature. This was the view a little more than a century ago, the view which was proved to be very wrong. Two revolutions in physics took place in the last century: relativity theory and quantum theory. Classical physics provided very plausible explanations of many observable phenomena, but it was shown to be wrong in more precise experiments.

 Today there are no paradoxes similar to those encountered by classical physics. However, the ontology of the quantum wave function seems very far from what we observe with our senses, and this is what led the majority to believe that it cannot be the true description of Nature. I, however, think that the classical understanding of Nature at the beginning of the last century is roughly correct also for quantum theory. An atom is not made of point particles moving on definite trajectories, but  is a highly entangled quantum wave of electrons, protons, etc. This quantum wave is, however,  not relevant for the interaction with other systems, it exerts forces on other systems very much like a point-like body. The wave structure provides an explanation of the spectrum, which classical mechanics does not, but for explaining pressure or other macroscopic features, classical approximations work fine.

In the following I will present the view that all that exists is the wave function of the Universe which evolves according to the Schr\"odinger equation and without collapses during quantum measurements. This is the Many-worlds interpretation (MWI) of quantum mechanics \cite{Everett}.

\section{Many worlds in a single Universe}

 In my view \cite{SEP} the MWI is  the only interpretation which  can be viewed as  a possible final theory of physics. It is a deterministic theory  without action at a distance as all physics we know today. It  is  (still) not widely accepted since it requires a radical change in our view of Nature: there are many worlds which exist in parallel at the same space and time as our own.

As a physical theory, the MWI is a theory about the Wave Function of the physical Universe. The equations are those of the standard quantum mechanical formalism: the Schr\"odinger equation or its relativistic generalizations. The theory postulates that the only ontology is this Wave Function which evolves in a unitary way, i.e., there is no collapse of the Wave Function. Since the collapse process is the only random element in physics and also the only source for action at a distance, randomness and action at a distance are eliminated from physics in the MWI.

Clearly, the most general and complete description of  Nature is not the Wave Function of all particles in the Universe. Such a picture does not describe the creation and annihilation of particles and field theory is required. In such a framework a wave functional describing the amplitude for the local fields  at each point in space replaces the  Wave Function. However, I do not foresee that this entails significant conceptual changes or difficulties.  Surprisingly, little effort has been made to look for a rigorous definition of such a functional. Appreciation of the success of the MWI in resolving the paradoxes of  standard quantum mechanics may lead to further research in this direction.

In spite of the name, I view the MWI as a theory of a single Wave Function of a single physical Universe. This is the only ontology. Observables, which are frequently considered as basic building blocks of the standard formalism (and which are the
basis of algebraic approaches to quantum theory) are not part of the ontology. Thus, the uncertainty relations of quantum observables do not lead to randomness, the theory remains deterministic since the Wave Function evolves deterministically \cite{qmdet}.

There are two parts to the MWI: physics and interpretation. Physics, the theory of the evolution of the Wave Function, is rigorously defined and tested up to the maximally possible precision. It is  a ``good'' theory: it has no paradoxes, it is complete, it  contradicts neither the spirit nor the letter of the special theory of relativity.  As a physical theory it is clearly   simpler than the spontaneous collapse theories \cite{Pearle,GRW},   Bohmian mechanics \cite{deBroglie,Bohm}, the consistent  histories approach \cite{Grif}, or underlying probabilistic theories \cite{Spek}.  Moreover, recent analyses of possibilities for alternative interpretations show that they cannot be more  parsimonious than the theory of the quantum Wave Function, see a recent experimental study \cite{Rigbauer}.

The difficult part of the MWI is the interpretation, the explanation of how the Wave Function of the Universe describes the world we observe in 3-space while the Wave Function ``lives'' in the configuration space of $3N$ dimensions ($N$ is the number of particles in the world).
The role of interpretation is much larger in the MWI than in other interpretations of quantum mechanics. In most other interpretations the connection between ontology, e.g. Bohmian trajectories, or values of observables, is simple, transparent and immediate, so frequently it is not discussed at all, while in the MWI, it is the main part to be analyzed. Interpretation  belongs to the realm of human science with different standards and methodology. It has a much wider range of acceptable approaches.  FAPP (for all practical purposes) \cite{Bell} definitions are good enough for discussing our experiences.

The main difficulty of connecting our experience with the ontology in the MWI is that it corresponds not just to our current experience, but also to a multitude of other experiences in parallel worlds. Clearly, there are very many parallel worlds now, although the number of worlds is not rigorously defined. However, even in the gedanken case in the framework of the MWI in which  the Wave Function now describes just one world in the Universe, the correspondence to our experience in this world is  not obvious. I will start with the analysis of such a single world.

\section{A World}

The concept of a world in the MWI does not belong to the physics part. It is not like a particle: the form of the Wave Function includes information about the number and the type of particles in the physical Universe. Since the Wave Function is all there is, it must also have information about worlds, but the definition of what a world means is not written in the Wave Function. It belongs to human science. A world is what a layman understands by the word ``world''. A quantum physicist might have a confusing concept of a world, he can think of a multiverse \cite{Deut}, about particles in a superposition, etc., so the concept might not be clear. A world in the MWI is defined as a world of classical physics:

\begin{quote}
A world is a collection of  objects in the Universe in definite states.
\end{quote}

In classical physics we can imagine that all particles have definite positions and velocities.
However,  the positions of every molecule of Earth's atmosphere are not really relevant for describing a world in the MWI.  It is enough to specify definite states of all macroscopic objects. Moving to the quantum domain, we do not even have a description in terms of positions and velocities of  particles.  In the standard approach, the notion of observables plays an important   role. Our experience is described through a set of values of the observables. An observer is aware only of measured observables and their measurements collapse  the Wave Function to  definite eigenstates  which provide the correspondence.

David Bohm, in support of his interpretation, noted that in every experiment we read the values of observables by observing   positions  in  space of the pointers of the measuring devices. Definite temperature corresponds to a particular position of a thermometer, etc. So, the world is specified by well localized positions of all macroscopic objects. The world is a concept in 3-space. It is a crucial point for explaining the connection between the quantum wave and our experience. In a theory with collapses at each quantum measurement there is a  single world in which the Wave Function is a product of wave functions  of all macroscopic objects in 3-space.  The classical picture of interaction according to which an object creates fields in 3-space which cause accelerations of other objects present in the location of the field is valid here. (Recently I found a local explanation of this form   for  an apparent counter example to this claim, the Aharonov-Bohm effect \cite{VAB}.)

Neuroscience does not have a clear explanation of human experiences. Considering experience as information flow allows to speculate that it might be defined in terms of spin wave function \cite{Albert92}. This can lead to some modification of the picture, but apparently will not change it dramatically and currently there are no signs that our brain operates with spins.  The picture I present here is that our experiences supervene upon the space distribution of matter, so the Wave Function which allows us to build a three-dimensional picture of particles in 3-space specifies human experiences.
The density of particles is gauge independent and also properly transforms between different Lorentz observers. Thus, the explanation of our experience is unaffected by the ``narratability failure'' problem \cite{Albert13}, the Wave Function description might be different for different Lorentz observers, but the local description remains the same.

In fact, in the standard approach with collapses at every quantum measurement ensuring a single world, the problem of correspondence between the ontology and our experience is not considered problematic. (The collapse process, however,  is a very serious problem for the physics part of the theory.) This connection between the collapsed Wave Function and our experience will be the basis of the connection with the experience in the framework of the MWI.

The exact Wave Function of a world is not rigorously defined. It is based on collapse, but there is no precise definition when exactly it happens. The collapse avoids superpositions of macroscopic objects in macroscopically different states, but the word ``macroscopic'' is a FAPP concept. Von Neumann proved that there is a very large flexibility where we put the cut between quantum and classical, i.e. when exactly the collapse takes place. In the collapsed wave we can see the classical world. The three-dimensional picture of  places where the wave density is large is the picture of the world we see around us. Instead of an {\it ad hoc} von Neumann collapse, we can consider the spontaneously collapsed Wave Function of the GRW-Pearle type theory. The quantum wave resides essentially in 3-dimensions. It is connected to our experience in a transparent way. The macroscopic parts of  objects collapse to  pure unentangled states  due to  strong interaction with the environment.  States of electrons in atoms, and atom states in molecules, etc.  remain entangled, but micro states of particles are not relevant to the description of a world.

The state of an electron in an atom is not relevant for the description of a world, but when we describe a quantum interference experiment  with single particles, their state is important. One way to avoid description of micro objects is to note that the macroscopic description of macroscopic preparation and detection devices in the experiment includes the information about these particles. I think that it is more convenient to add a description of such micro particles to the picture of the world since it provides a clear characterization of the weak coupling of the particles with the environment. However, such micro particles require a special treatment, in particular, adding a backward evolving wave function specified by the final measurement \cite{tisy,Vpast}.

Let me summarize the main points. In a world, by definition, all macroscopic objects have definite macroscopic states.  In a world, there cannot be a Schr\"odinger's cat  in a closed box in a superposition of being alive and dead.
I add to the mathematical formalism of the Universal Wave Function a postulate regarding the connection to our experience. The connection between experience and the world wave function which is an element of the Universal Wave Function is the same as in the collapse theories in which there is only one world wave function and single experience of every observer. The Wave Function of a world $|\Psi_{\rm world}\rangle$ is a product state of well localized  (in 3-space) quantum states of all ``macroscopic'' objects multiplied by possibly entangled states of microscopic systems irrelevant for the macroscopic description of the world.

\section{The Wave Function of the Universe}

The correspondence between the Wave Function of a single world and our experience may not be as transparent as the correspondence between our experience and a classical world of particles moving on trajectories, or a world of Bohmian trajectories, but I still find it satisfactory. I do not believe that this is the picture of the Universe because there is no satisfactory physics which leads to such  Wave Function. The collapse process which is required here is foreign to all physics that I know. It has some kind of action at a distance and randomness and nobody has found elegant and simple equations for the collapse process. Physics forces me to reject the idea of collapse, and therefore, after every quantum measurement a superposition of wave functions of the single world wave function type is created:
\begin{equation}\label{Univ}
| \Psi_{\rm Universe}\rangle=\sum \alpha_i | \Psi_{{\rm world}~i}\rangle.
\end{equation}

The Wave Funciton of the Universe is a wave function in a configuration space $|\Psi_{\rm Universe} (x_1, x_2,...x_N)\rangle$. Each one of the wave functions of the worlds  is essentially a product of well localized wave functions of macroscopic objects in 3-space $|\Psi_{{\rm world}~i}\rangle= \prod_j |\Phi_i(X_j)\rangle\ |\psi_i\rangle$, where $|\psi_i\rangle$ signifies the wave function of other degrees of freedom: relative coordinates of electrons in atoms, weakly coupled particles like neutrinos, etc. Locality and strength of  interactions (with a buzz word ``decoherence'') ensure approximate uniqueness of the decomposition.

Different worlds must have  different classical descriptions and therefore they correspond to orthogonal wave functions. Energies are   bounded, so even  well localized wave functions of macroscopic objects must have nonvanishing tails at macroscopic distances. This makes the above statement of orthogonality not so obvious. Most probably, the world wave functions of different worlds can be constructed to be orthogonal, but even if not, since the concept of a world is a human FAPP concept, a tiny overlap can and should be neglected.

The  Wave Function of the Universe $| \Psi_{\rm Universe}\rangle$ can be decomposed into a superposition of the world wave functions $| \Psi_{{\rm world}~i}\rangle$. The reason we do not experience superpositions is not because they do not exist, but because {\it we} are not capable of experiencing  several different states simultaneously. The phrase ``different states'' means different places and the locality of physical interactions prevents conscious nonlocal creatures. A nonlocal creature which behaves differently depending on the relative phase $\phi$  between parts at different locations, $\frac{1}{\sqrt 2} (|A\rangle + e^{i\phi} |B\rangle)$, cannot be useful because local interactions wash out the relative phase almost immediately. This is the argument for the ``preferred basis'' of the decomposition of the Wave Function, it has to be the local basis.

Another argument for why the Universe cannot support worlds with nonlocal   creatures being in a superposition of two locations is that such a creature  will need much more resources. If such a nonlocal creature would like to eat, it will need two dinners instead of one. Both parts of the creature $ |A\rangle$ and $|B\rangle$ have to eat, but a superposition of a dinner in two locations $\frac{1}{\sqrt 2} (|A_{\rm food}\rangle + |B_{\rm food}\rangle)$ cannot provide a dinner for the creature through local interactions. Such interactions cannot transform a product state of the creature and the food into the required state $\frac{1}{\sqrt 2} (|A\rangle|A_{\rm food}\rangle  + e^{i\phi} |B\rangle|B_{\rm food}\rangle)$.

The {\it stability} of  world wave functions $| \Psi_{{\rm world}~i}\rangle$  is the key issue.  Of course, at every quantum measurement the world wave function splits to several different (well localized) world wave functions. But this does not happen too fast and too often. If no observer has the time to be aware of a particular world, we have no reason to define such a world: remember, `world' is a human concept which is supposed to help explaining our experience.
I do want to consider worlds at a particular time, but it is not enough that the picture drawn from a wave function looks like a world  (all macroscopic objects are well localized) at this moment. If, immediately after, it does not look like a world, such wave function should not be considered as corresponding to a world.

We can ask what was the past and what will be the future of a world.  Every quantum measurement splits the world, so a world in a particular time corresponds to a multitude of worlds in the future. Going backward in time we have a unique past. For many worlds at present there is the same past. So, if we consider worlds for periods of time we obtain ``overlapping'' worlds: they overlap in the past. (If we want to add some micro particles for describing the world, then splitting for them happens at the measurement {\it after} the present time of the world we consider \cite{tisy}.)

Another important concept which has to be clarified is ``I''. In which world am I? If I perform a quantum measurement what happens to me? If somebody else performs a quantum measurement will I be changed?

By definition, I have to be in a well defined state. All parts of my body, neurons in my brain, etc. have to be well localized. When I perform a quantum measurement with a few possible results and I observe an outcome, a few worlds are created with a different ``I'' in each world. All these ``I''s are descendants of the ``I'' before the measurement, so they have identical memories of what happened before the quantum experiment, but they have different knowledge about the result of the experiment.  ``I''s which have identical memories, but different locations of bodies, are also considered different. Such situations can be created when I am moved to different places, while asleep, according to the results  of a quantum experiment  \cite{schizo}).

If somebody else makes a quantum measurement in a faraway location such that the information about the result does not arrive at my location, then I do not split. The world splits, but I remain part of all newly created worlds. There is no meaning for the question what is the result of this measurement in {\it my} world. (This is one of the examples in which our language, developed during the time of a firm belief in existence of a single world, has difficulty in describing the situation.) So, it cannot be that different ``I''s are present in one world, but it is possible that the same ``I'' is present in several worlds.

Let me summarize. I build the MWI on the basis of standard quantum mechanics. I ask physicists who accept Wave Function collapse to provide (a rough) description of the Wave Function of the physical Universe. This specifies what might be a world wave function. Next,  I decompose the Wave Function of the Universe as a superposition of such world wave functions. I postulate the correspondence with the experience as the one in   standard quantum mechanics. Thus, the presence of a term corresponding to a world $i$ in the decomposition, $\alpha_i\neq 0$, ensures the presence of such an experience. It is a question for  quantum physicists what should be the world wave function to ensure stability and ``classicality'' of this world.

 The wave function of a world $|\Psi_{{\rm world}~i}\rangle$ specifies what we  ``feel'' in world $i$. What is the role of the coefficient $\alpha_i$? The absolute value of the coefficient  specifies the illusion of probability as will be discussed in the next section. The phase is not relevant to us. We assume that worlds are different enough that we cannot make interference experiments between them. The worlds differ by macroscopic differences of states of macroscopic objects. The FAPP meaning of the word ``macroscopic'' can be defined exactly by this property: the states are macroscopically different if we cannot observe interference between them in a realistic experiment. The phase of $\alpha_i$ can be relevant for a creature having unlimited technological power to perform interference experiments with macroscopic objects, i.e., interference between different worlds.

 \section{Measure of existence of the world and the illusion of probability}

 The measure of existence of the world, $\mu_i=|\alpha_i|^2$ ,  provides the illusion of probability in our world. Before discussing the qualitative issue, why $ |\alpha_i|^2$ and not something else, I will discuss the meaning of probability in the MWI.

 I write ``illusion'' of probability, because I consider a genuine concept of probability to require that there are several options and only one takes place. In case of quantum measurements all possible outcomes take place.

There is no randomness in the basic laws of physics. But determinism by itself does not prevent the concept of probability. We might be ignorant of some details which specify the outcome. In quantum mechanics there are no such details (which usually are named ``hidden variables''). We might have complete information about the system and the measuring device and it still will not help us to know what will be the outcome. More precisely, we know that all outcomes will take place and we say that each outcome will correspond to one of the newly created worlds.

If there is a collapse of the Wave Function to one of the world wave functions $|\Psi_{{\rm world}~i}\rangle$, then we do have a legitimate concept of probability. An experience of an observer supervenes upon the world wave function. In the MWI, every term in the decomposition (\ref{Univ}) corresponds to such experience. By construction, our experience in each particular world is the same as our experience in a physical universe with the collapse law in which  only one world exists. So, we have a complete illusion of probability: there is no difference between our experience and the experience of an agent with genuine probability.

Since the MWI is a deterministic theory, there cannot be a genuine chance there, but we can arrange a special situation in which there will be  a genuine ignorance probability of an agent about the outcome of a quantum experiment. Before, I pointed out that in quantum mechanics we can know everything about the past without knowing what will be an outcome in the future.
 Now I will show that in the MWI we might know everything about the present, but still be ignorant about the outcome of already performed experiment.

 Such a situation is not easy to understand  because we are not used to considering a plurality of worlds. What might help is to imagine first a gedanken situation which is not related to quantum physics. Suppose that   creatures with super technology land on Earth. They can do what we would consider to be miracles. The creatures can create copies of Earth with everything on it and add it to the Solar system. They show their ability to people on Earth and they say that tomorrow morning there will be three identical Earth planets with all  inhabitants. In the morning I wake up and I do not know if I am the original, or I am one of the two copies. Three Earth worlds exist and there are three ``I''s. Each one does not know which ``I'' he is. The symmetry tells us that the probability to be each particular ``I'' is one third.  Each ``I'' will bet one third on the fact that he is the original and not a copy.

 In the case of a quantum measurement I can arrange a similar situation without super technology \cite{schizo}. Before performing a quantum experiment which has three possible outcomes I take a sleeping pill and ask my friends to move me while I am asleep to one of three rooms according to the result of the experiment. My friends, instead of performing the experiment themselves, can  get instructions using the iPhone application ``Universe Splitter''  or the Tel Aviv World Splitter \cite{TUWS}. When ``I'' in a particular room am  awake, but still have closed eyes, I  do not know in which room  am I. It is an unusual situation: I, more precisely, every copy of me, might know everything about the Universe, but still be ignorant about the outcome. I am ignorant about self-location in a particular world. This is my privileged property.   Any external observer does not have this probability concept. The question: ``In which world am I present?'' has no meaning for him. There are different ``I''s present in corresponding different worlds, but the external observer,   even if he is aware that the splitting has occurred,  belongs to all these worlds. Only when he contacts me, will he split his ``I'' according to the outcome of the quantum experiment.

 A widespread  approach to probability is the readiness to put  an intelligent bet on a particular outcome. Since the probability is relevant for all my descendants  as they    have a legitimate concept of ignorance probability, I suggest to associate it also with me at the time before the experiment. It is rational for me to place bets since all my descendants would like me to do so. They will get the rewards (and the losses) of the bet. Thus, considering the probability as an amount that an intelligent agent is ready to bet on a particular outcome \cite{deFin}, we have a concept of probability in the framework of the MWI. The rule is that an experimentalist performing a quantum experiment should bet in proportion to the {\it measure of existence} of the world with this outcome \cite{SBAna}.

 In classical cases either a world exists or it does not. In case of several planet Earths, there is no way that (at a particular time) one planet exists more than the other. We must associate the same measure of existence with all planets: just 1.   The definition of the measure of existence of a quantum world, $\mu_i=|\alpha_i|^2$, allows different values. I postulate, that we should  bet according to this measure. This analog of the Born Rule  in the framework of the MWI is sometimes named the Born-Vaidman rule \cite{Tap}.

  Why accept this postulate? In my view, a good answer is that the records of the outcomes of quantum experiments in our world show that there is no general rule which   leads to   better results for an experimentalist betting on the outcomes of quantum experiments. One can add that the postulate is natural if we accept the idea of measure of existence. Indeed, arranging (artificially) that all worlds have equal measures of existence leads to a natural property of equal probability for all worlds as in the case of  multiple ``classical'' worlds (compare with my three Earths example). The measure of existence of a world  also quantifies the power of a world to interfere with other worlds if a quantum super technology will make interference experiments with macroscopic objects which are interference experiments between different worlds \cite{schizo}.

 Another argument for accepting the Born-Vaidman Rule is that all the alternatives I try to imagine lead to contradiction with special relativity. If we have a wave function of a particle distributed in space, the sum of the probabilities to find it in some location adds to 1. Thus, if we can change the probability in one place  by some local operation, this will change the probability in another place. Changed probability is a signal, since a large ensemble of identical systems with identical actions allows transmission of a message. Local operations at a local part $O$ can change almost all local aspects of the wave function  except for changing its weight, $\int_O\Psi^*\Psi dv$. So, it seems to be that the probability cannot depend on any other function of the wave function but its local weight.

  \section{Conclusions}

The MWI, first, tells us what the ontology is: it is the evolving Wave Function. Then it finds us in this ontology, as stable waves in human shapes, and explains  (through some natural postulates) our experience.

 The theory of the evolution of the Wave Function is an exact mathematical theory of a single physical Universe.
  Avoiding discussion of values of variables makes it free of paradoxes, action at a distance, and randomness, the features which are foreign to all known areas of physics.

The second part of the MWI, the interpretation, is good only as a FAPP theory. It allows  various modifications, e.g., the original Everett relative state formulations provides a somewhat different decomposition to world wave functions.

The MWI is far from being the most clear  explanation of our experience and although it can probably  be improved, some other interpretations certainly explain it much better. However, these interpretations pay a very high price in spoiling the physical part of the theory which becomes so unnatural  that most physicists are not ready to accept them.

My impression is that the majority of physicists are not ready to accept the MWI because in this framework {\it we}  are not in the center of the theory. The sun does not encircle us. Physicists are still hoping that there will be some new theory (interpretation) which will be better than the MWI. I doubt it, but I think that it should be at least accepted that meanwhile the MWI is the best alternative we have today.

This work has been supported in part by the  Israel Science Foundation  Grant No. 1311/14.


\end{document}